\documentclass[review]{elsarticle}

\usepackage{subcaption}

\journal{European Journal of Mechanics - B/Fluids}

\begin{document}

\begin{frontmatter}

\title{\textbf{ Measuring surface gravity waves
using a Kinect sensor}}

\author[address_unito]{Francesco Toselli\corref{mycorrespondingauthor}}
\author[address_unito]{Filippo De Lillo}
\author[address_unito]{Miguel Onorato}
\author[address_unito]{Guido Boffetta}

\cortext[mycorrespondingauthor]{Corresponding author}

\address[address_unito]{Dipartimento di Fisica, Universit\`{a} di Torino and INFN, Sezione di Torino, \\
Via P. Giuria, 1 - Torino, 10125, Italy}

\begin{abstract}
We present a technique for measuring  the
two-dimensional surface water wave elevation both in space and time
based on the low-cost Microsoft Kinect sensor. We discuss the capabilities 
of the system and a method for its calibration.
We illustrate the application of the Kinect to an experiment in a small wave tank. A detailed comparison with standard capacitive wave gauges is also performed. Spectral analysis of a random-forced wave field is used to obtain the dispersion relation of water waves, demonstrating the potentialities of the setup for the investigation of the statistical properties of surface waves.
\end{abstract}

\begin{keyword}
 Surface gravity wave\sep Space-time measurements \sep Dispersion relation
\end{keyword}

\end{frontmatter}

\section{Introduction\label{intro}}
Surface gravity waves are waves that propagate on the interface between two
fluids, for example air and water, in the presence of a gravitational force
that acts as a restoring force.  One of the most common examples is given by
ocean waves that can propagate for hundreds of kilometers, after being
generated and forced by the wind. Their dynamics is ruled by the Navier-Stokes
equations; however, due to the complexity of such equations one usually relies
on simplified models or on experiments. Traditionally, the displacement of the
free surface  from its equilibrium position has been measured using floating
buoys (see for example \cite{nationalocean,kinsman1965wind}): their vertical
acceleration can be used, after double integration, to measure the surface
elevation in a single point as a function of time. The same kind of measurement
can be performed using capacitive or resistive wave gauges; because of their
fragility those are usually adopted in more controllable environments, even
though measurements in open seas have been performed (see for example
\cite{cavaleri1987reynolds} or more recently \cite{resio2004equilibrium}). In
the pioneering work in \cite{longuet1963observations}, an account of a method
for estimating the directional wave spectra has been given: the idea relies on
the measurement not only of the vertical acceleration of a buoy but also of the
angle of roll and pitch. Using those observables one can extract a limited
number of Fourier coefficients of the angular distribution of the energy in
each frequency band, reconstructing an energy spectrum as a function of
frequency and angle of propagation of the waves.  Despite the extremely
important role played  by such measurements for the understanding of many
properties of ocean surface gravity waves (including spectra and other
statistical quantities), those techniques are not satisfactory if one is
interested in addressing the problem of the dynamics of surface  waves; indeed,
starting from the late fifties (see the pioneering work in
\cite{cox1958measurements}), optical techniques have been developed for
application to ocean waves. Images of the surface elevation provide an
indication of the spatial properties of the waves and can be used to compute
wave number spectra, see \cite{hwang2000airborne1,hwang2000airborne2} where an
airborne scanning lidar system as been developed to measure spatially ocean
waves; however, what is really relevant for studying the dynamics of the waves
is a sequence of images so that waves are measured both in space (2D) and time.
A recent important development in this direction has been provided in
\cite{benetazzo2006measurements,benetazzo2012offshore} (see also
\cite{klette1998three}) where stereographic recordings of ocean waves have been
performed. Such measurement techniques allow one to reconstruct the dispersion
relation of water waves, \cite{peureux2018note}. Optical methods based on
profilometry, \cite{cobelli2009global,cobelli2011different} have also been
developed to measure the 2D surface elevation in space and time and extract
different regimes of wave turbulence in small scale experiments. 
\\Despite these developments of the last 15 years, simple, accurate and inexpensive techniques for measuring surface gravity waves in space and time are matter of research. 

We have developed a high resolution method to measure the 2D elevation of water waves based on the Microsoft Kinect. The main purpose of this study is to evaluate the possibility to use the Kinect depth sensor to reconstruct the time-dependent configuration of surface waves of moderate amplitude.

The remaining of this paper is organized as follows. Section~\ref{methods} is 
devoted to the discussion of the hardware and the software and the
calibration procedure. In Section~\ref{results} we discuss the results of the measures performed on water waves produced in a lab setup. The pointwise wave height measured with the Kinect is compared with measurements obtained by capacitive probes and the wave-fields obtained with the Kinect are used to reconstruct the dispersion relation of gravity waves. Finally, in Section~\ref{conclusions} conclusions and perspectives will be discussed.

\section{Methods\label{methods}}

The core of the acquisition system is the Microsoft {\it Kinect} (Windows edition) \cite{kinectMicrosoft},
 which includes a depth sensor based on  infrared (IR) imaging.
Producing data similar to a light detection and ranging sensor, the Kinect has
the advantage to be cheaper and smaller than other tools commonly used to
assimilate this kind of data. These characteristics make it a good candidate
for practical use in providing high-level precision measurements in Earth
science studies such as water waves amplitude reconstruction, and free-surface flow and bathymetry measurements \cite{mankoff2013kinect}. 

\subsection{Hardware\label{hardware}}

The Kinect sensor contains an IR emitter coupled with an IR depth sensor camera
and range images are produced from the so-called light coding technique. The IR
emitter generates a pattern on the object under study: this object needs to be
opaque in order to diffuse the projected pattern which will be captured by the
IR sensor. The image received by the sensor is compared with the original
pattern by an on-board processor which uses the relative distortion to
associate depth information to each pixel. The Kinect returns the depth field
with a high spatial resolution of $640 \times 480$ pixels and 11-bit depth
quantization. It operates with a refresh rate of $30$ Hz and, depending on the
recording mode chosen, it  can be used in a "default" (nominal optimal depth
range between $1$ - $6$ m) or a "near" mode ($0.5$ - $3$ m).  In studying the
capability of the sensor to detect wave amplitudes, we found an optimal working
distance between $700$ mm and $800$ mm using the device in near mode.  In this
range the horizontal/vertical and depth resolutions of the sensor is around $1$
mm \cite{viager2011}.  The Kinect IR camera has a nominal field of view (FoV)
of about $58 \times 45$ degrees, therefore the dimension of the recorded region
placed at $750$ mm from the sensor is about $800 \times 600$ mm$^2$.
\\Together with the depth sensor, the Kinect is equipped with an RGB camera at
resolution $1280 \times 960$ pixels which can acquire simultaneously with the
IR sensor and which is therefore useful to check the recorded area, a
4-microphone array and an on-board accelerometer which can be used to measure
the device orientation in space. The simultaneous use of the RGB and IR
cameras requires specific calibration in order to obtain the correct
correspondence between the respective FoV. However, we did not use the RGB
sensor, instead inferring all geometrical information from the IR camera.  

\subsection{Software\label{software}}

The software interface was composed by a first module that acquires the data
from the sensor and stores it on disk and a post-processing module that performs
the actual wave detection. The first module, implemented in C\#, leverages the
Microsoft API (Kinect for Windows SDK 2.0) to interact with the device driver, and includes a graphical
interface to help position the sensor.
Afterwards the sensor data is stored as an image in the flexible PNG format,
 enabling loss-less compression and high channel depth (in our case, 16 bit).

\subsection{Calibration of the sensor\label{calibrations}}
\begin{figure}[h!]
\centering
\includegraphics[width=.8\linewidth]{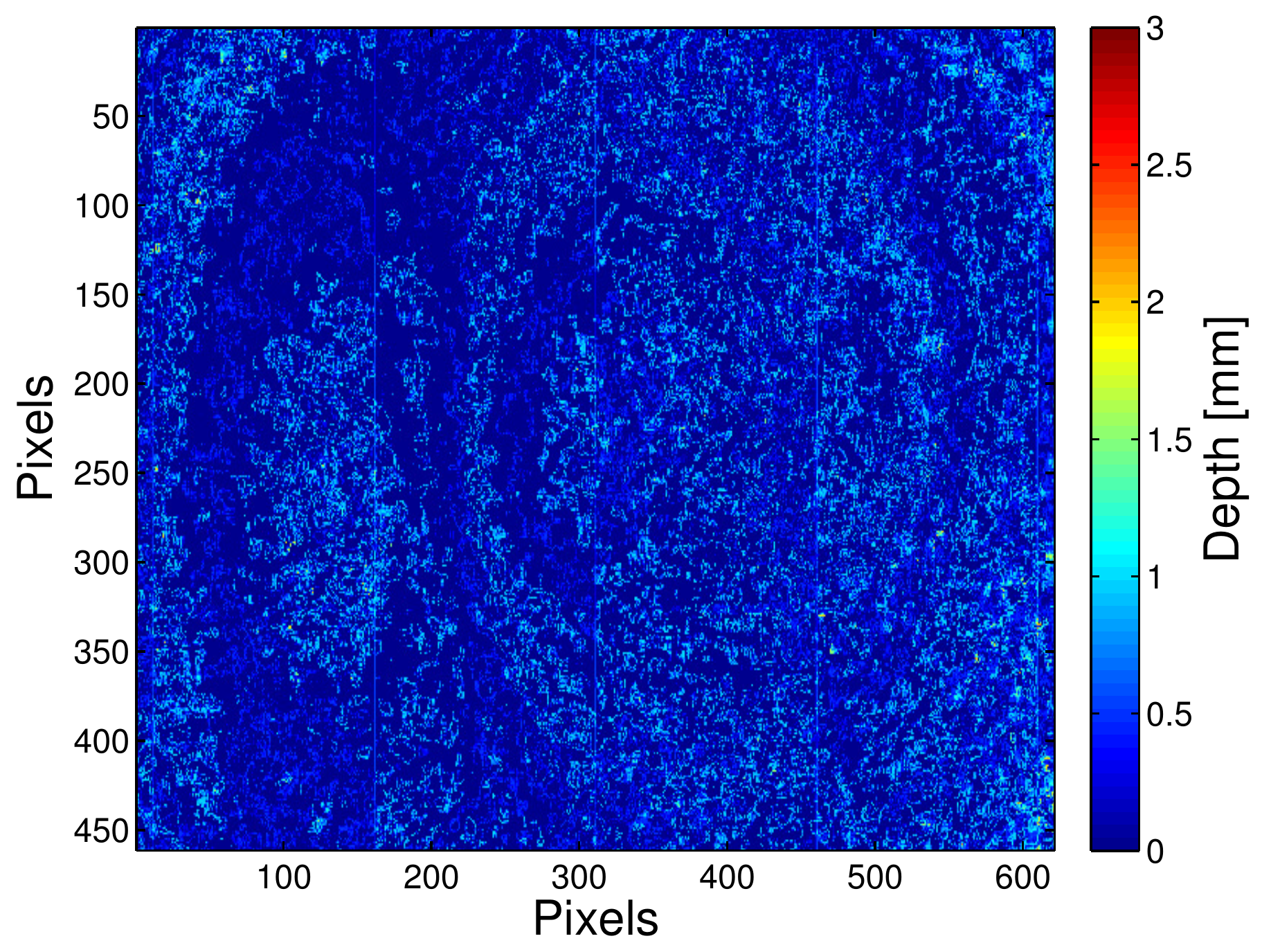}
\caption{Two-dimensional standard deviation map computed by using 
$3$ s of record from a flat fixed surface. In most of the pixels 
the standard deviation is of the order of the nominal vertical 
sensibility of $1$ mm.}
\label{fig1}
\end{figure}

The calibration of the sensor was performed in several steps.
A flat plane was positioned horizontally with a
spirit level with a precision of $0.1$ degrees. The sensor was installed on a
3-axis camera mount by using the values acquired by its internal accelerometer. These values were taken as references for the following placement of the sensor in the actual experimental setup. 
\\We measured the noise level of the depth signal when viewing a flat surface
at fixed distance. Time sequences of 90 images were acquired and the standard deviation $\sigma_d$ of the depth of each pixel was computed.
The resulting map of standard deviations 
is shown in Fig.~\ref{fig1}. For most of the pixels $\sigma_d$ is lower than the nominal vertical resolution of $1$ mm. 
\\We have then calibrated the depth sensor by collecting data for
a flat surface placed at $11$ different known distances from the sensor 
in the range $600-900$ mm. For each distance one minute of data ($1800$
images) are acquired and these depth fields are averaged together to eliminate noise.
For each pixel $(px,py)$ of the averaged image
we use a linear model to connect the depth value $pz$ to the 
physical distance $z$ as
\begin{equation}
z=m(px,py) pz + q(px,py) 
\label{eq1}
\end{equation}
which defines the two matrices of coefficients $m(px,py)$ and $q(px,py)$.
We found that the linear model works very well with global 
regression coefficient $R=0.998$. 
We remark that the model \ref{eq1} neglects possible aberrations due to the lens. This assumption was verified by systematically analyzing the position of the digital image of objects of known physical position (spanning both the horizontal and vertical ranges of the instrument), which gave no measurable distortion within the resolution of the depth sensor.

\subsection{Experimental setup}
\begin{figure}[h!]
\centering
\includegraphics[width=.8\linewidth]{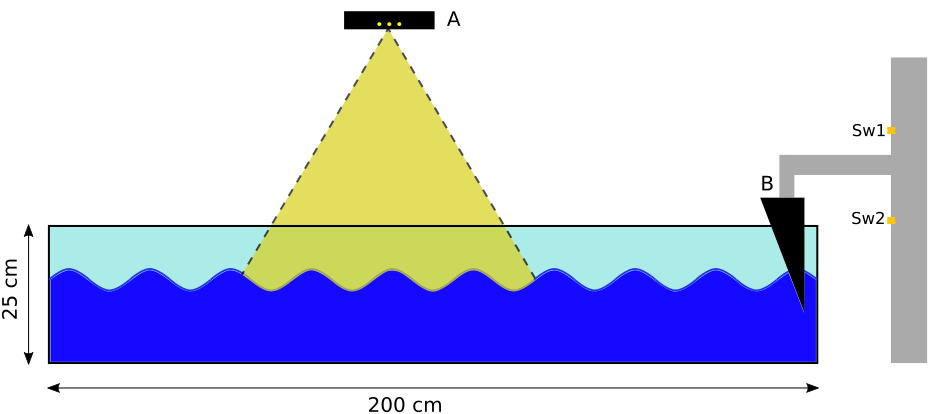}
\caption{Scheme of the experimental setup (not to scale). The FoV of the sensor A is positioned several wavelengths away from the wavemaker in order to avoid spurious fronts coming from the wedge B. The linear motor inverts the motion of wedge when it reaches one of the switches Sw1 and Sw2.}
\label{fig2}
\end{figure}

In order to verify the possibility to measure the dynamic evolution of a surface-wave field, we placed the sensor over a wave tank and performed recordings of the evolving surface. Figure~\ref{fig2} shows a schematic representation of the setup.
We use a small linear wave tank, of horizontal sizes $2000 {\rm mm} \times 500 {\rm mm}$ and depth $250$ mm.
\\A wave generator was placed on the short side. No absorber was used on the opposite
side, allowing for standing waves. 
The wave generator consists of a wedge fixed to a vertical motorized linear guide. The wedge moves between two switches which invert the linear motor. The motion of the wedge in the water forces waves whose amplitude and frequency is controlled via
the positions of the switches (changing the depth of immersion) and the speed of the linear motor.
\\As discussed in \ref{hardware}, the Kinect cannot detect the surface of pure water because of its transparency. For this reason the tank was filled with about $18$ cm of water with the addition of white liquid dye: this method allowed to make the liquid opaque and detect the water surface \cite{combes2011free}.
\\The Kinect was installed on the 3-axis camera mount used for the calibration
at a distance of about $75$ cm from the fluid surface. The orientation of the sensor was set consistently with the calibration setup using as reference the readings of the internal accelerometer.
\\Kinect acquisitions were complemented with local measurements
of surface height, obtained with a wave gauge system (INSEAN, Rome) consisting of two capacitive probes read by an interface which digitizes the signal and connects with a personal computer. The system has a 
a sensitivity of 1 mm and the probes were placed inside the tank
just outside the Kinect FoV, in order not to interfere with the images acquired.

\section{Results\label{results}}
\begin{figure}[h!]
\centering
\begin{minipage}[c]{0.8\columnwidth}
\includegraphics[width=1.0\linewidth]{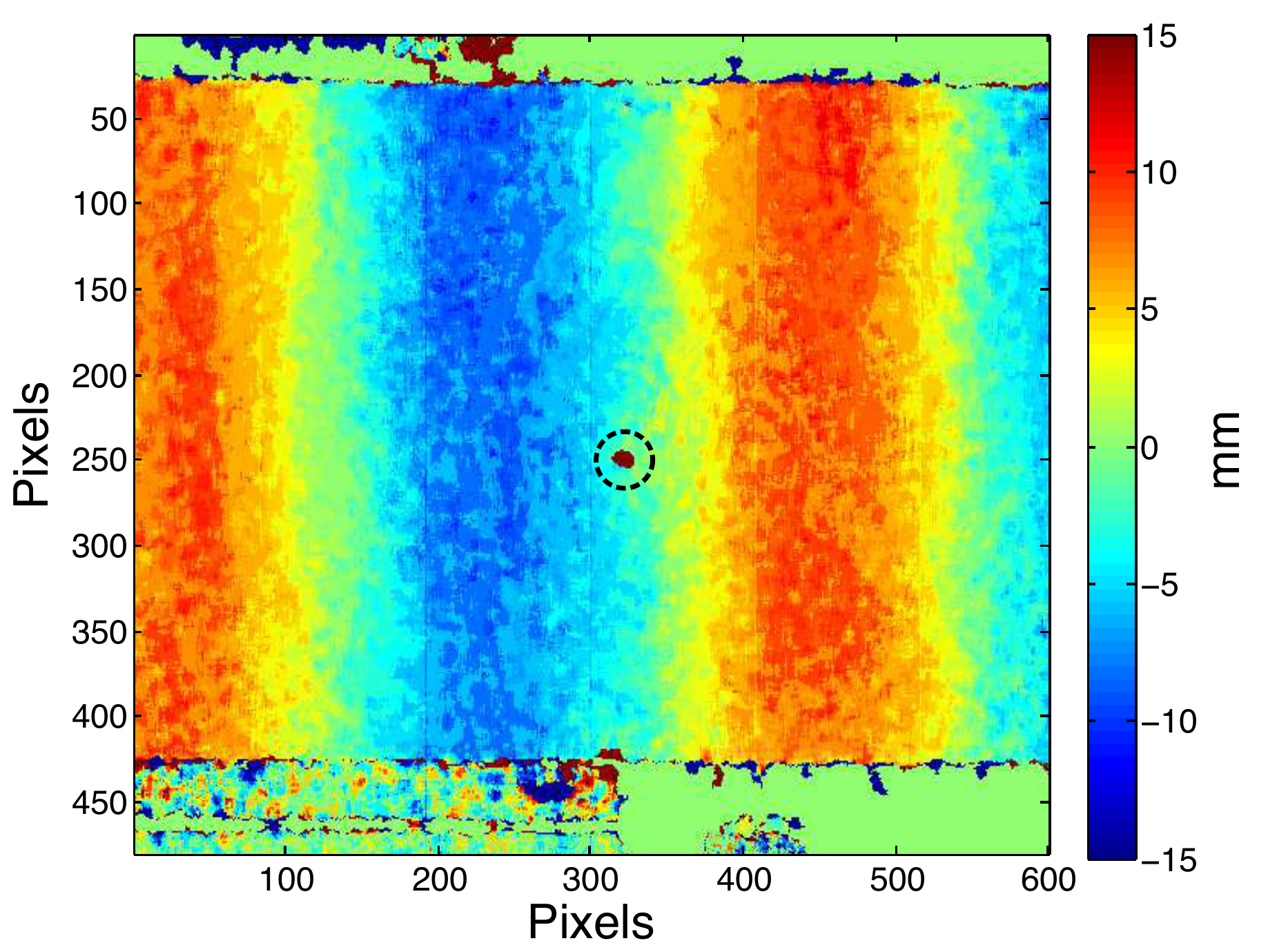}
\put(-270,190){(a)}
\end{minipage} 
\begin{minipage}[c]{0.8\columnwidth}
\includegraphics[width=1.0\linewidth]{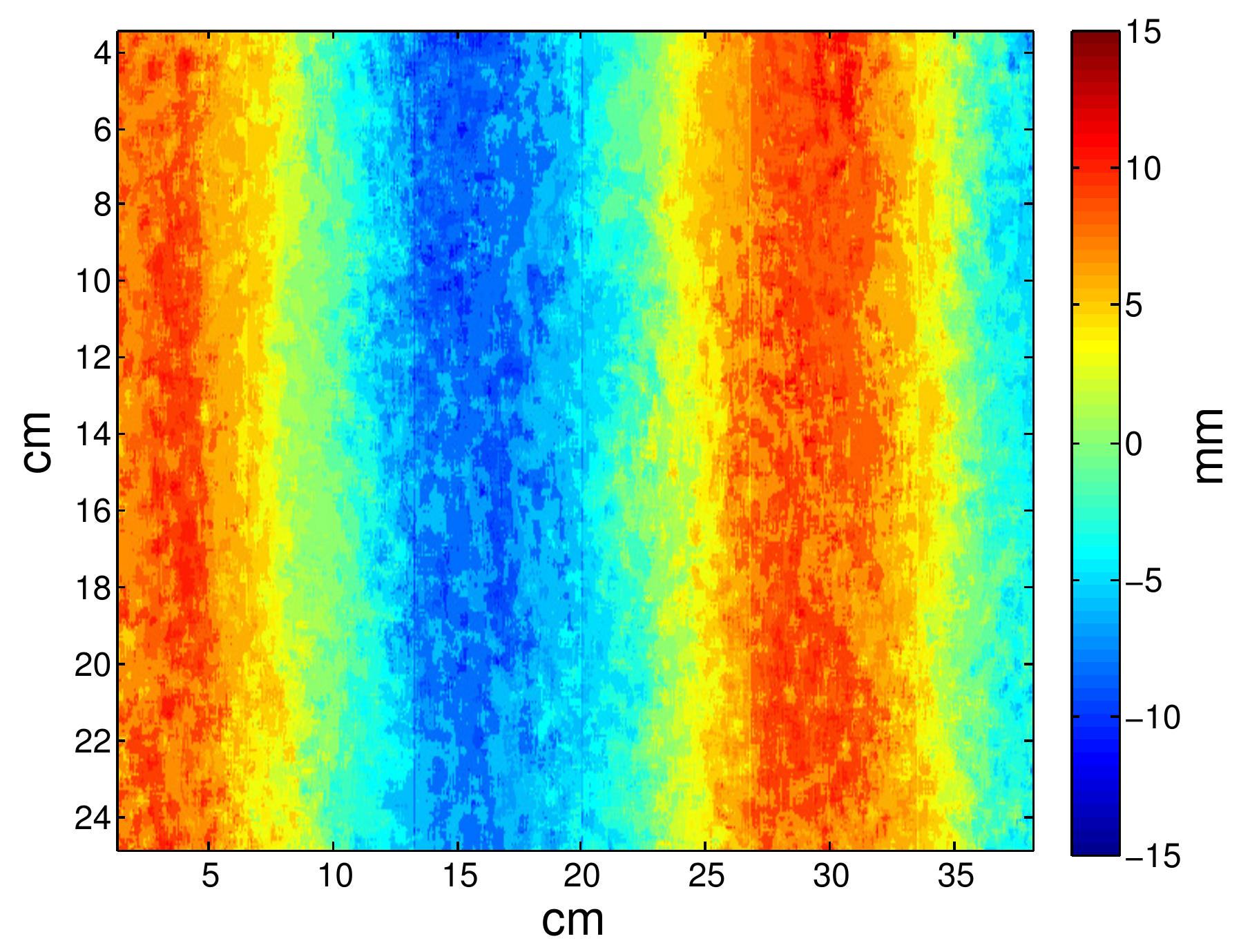}
\put(-270,190){(b)}
\end{minipage} 
\caption{Depth field before and after post processing. (a) Original Kinect depth data: regions outside the tank can be identified, as well as a limited number of pixels in the region inside the dashed black circle corresponding to the central area of the FoV where direct reflection produces spurious distances. (b) Depth data after cropping and removing reflective spots with interpolation.}
\label{fig3}
\end{figure}
\begin{figure}[h!]
\centering
\begin{minipage}[c]{0.75\columnwidth}
\includegraphics[width=1.0\linewidth]{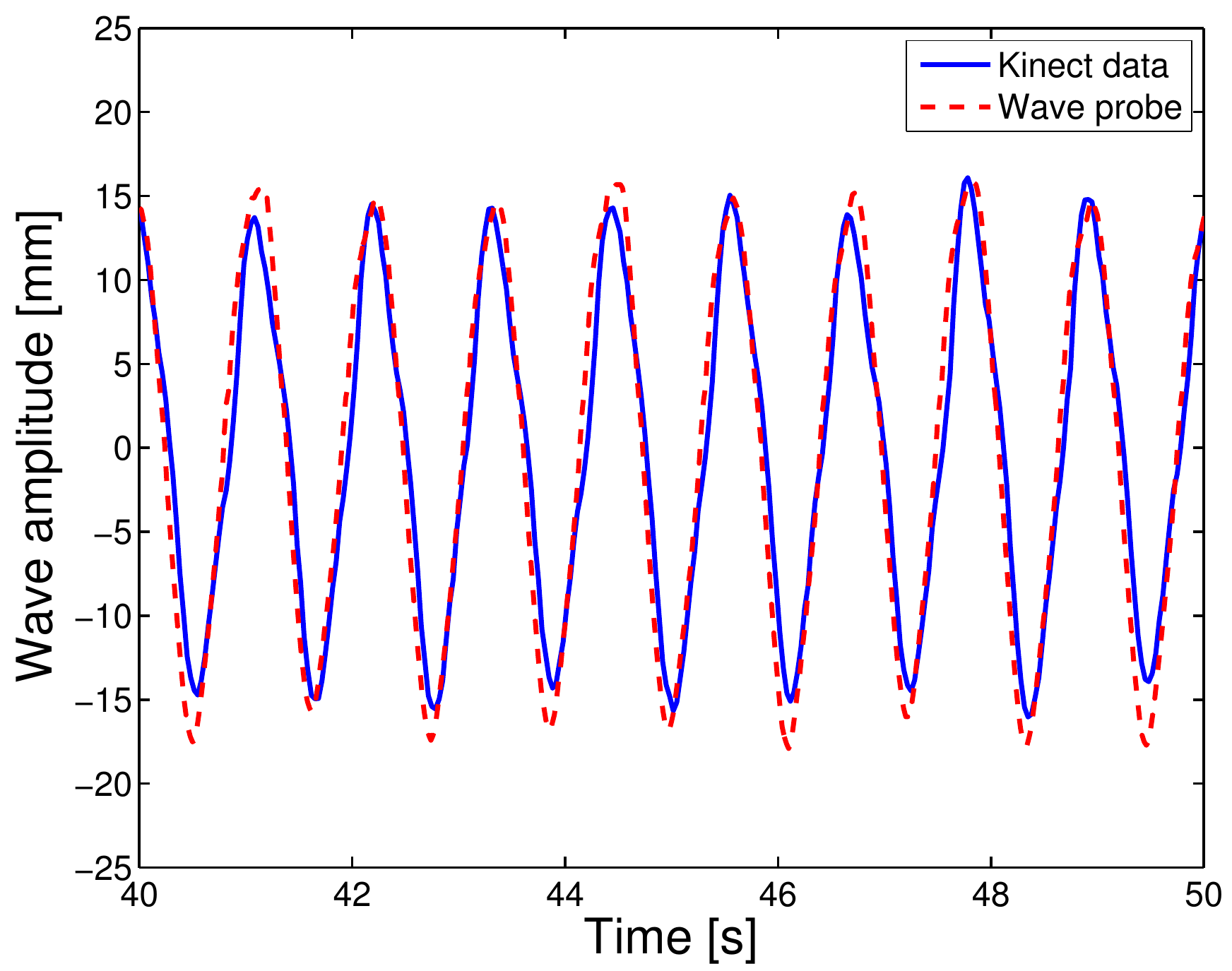}
\put(-260,190){(a)}
\end{minipage} 
\begin{minipage}[c]{0.75\columnwidth}
\includegraphics[width=1.0\linewidth]{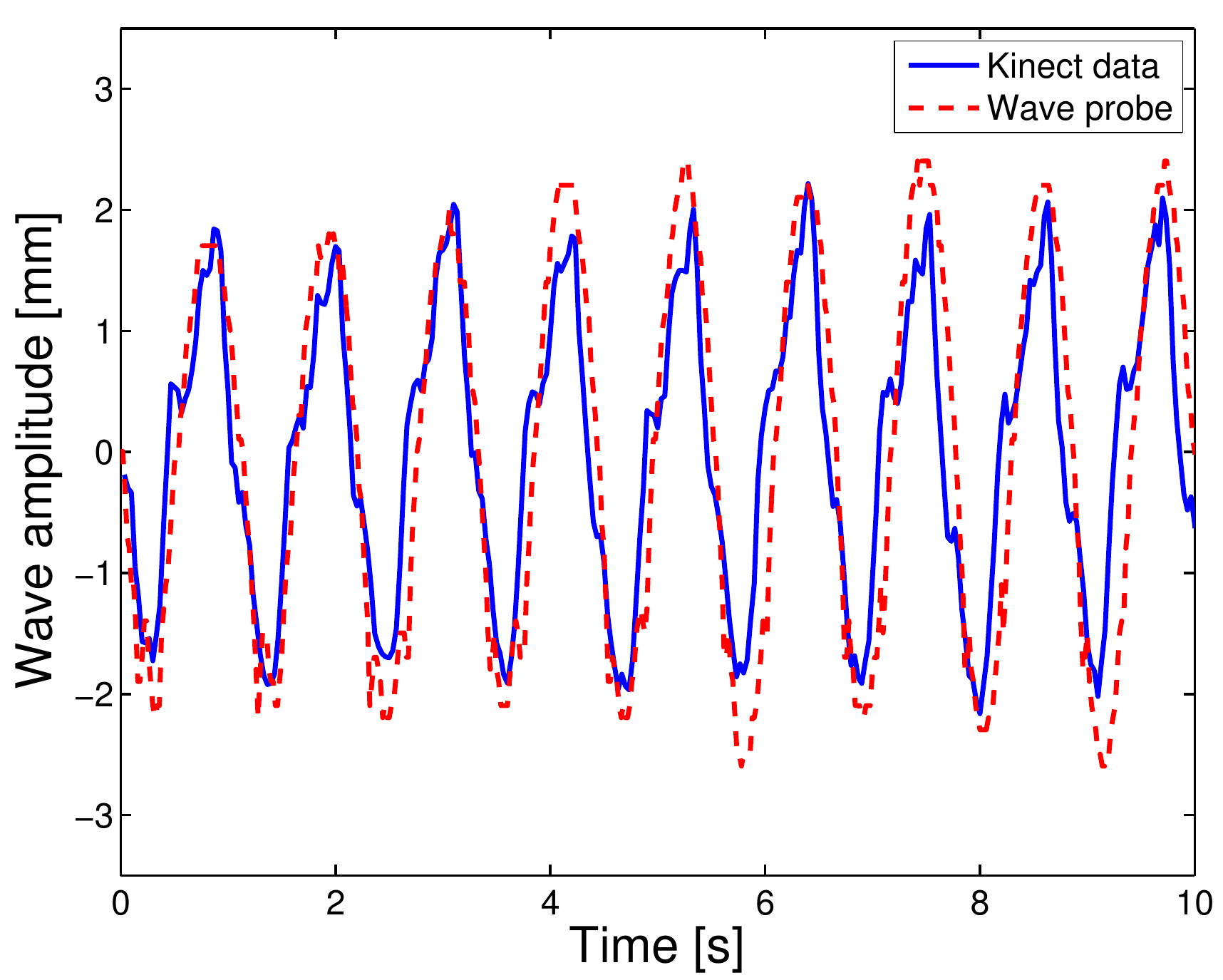}
\put(-260,190){(b)}
\end{minipage} 
\caption{Time series of the depth signal averaged over a line of pixels at the same longitudinal position, near (and parallel to) the edge of the FoV closest to the wave-maker. Panels (a) and (b) were obtained with the same forcing frequency and different amplitude. The Kinect data (solid line) are compared with the measurements from a capacitive probe (dashed line) a few centimeters away, just outside the Kinect sensor FoV. The signal for the smaller wave (b) is clearly more noisy, but it is for both instruments and still shows good agreement between the two methods.}
\label{fig4}
\end{figure}

By changing the amplitude and the frequency of the wave forcing, we could test the potentiality of our setup for the analysis of wave fields of amplitude between $2$ and $20$ mm.
\\As described in the previous sections, at each time frame the sensor outputs a 2D image where the amplitude of each pixel gives the distance of the corresponding point in space from the device. Apart from the intrinsic noise in the measure, some limitations are specific to the method of measurement. Of particular relevance for experimental fluid mechanics is the presence of some reflection of the liquid surface. Indeed, in order for the device to correctly infer the distance of surfaces, a clean, purely diffusive image of the projected pattern must reach the IR sensor. 
As a consequence, the Kinect is unable to provide depth data for some pixels (see Fig.
\ref{fig3}) due to direct reflection of the IR pattern at small
angles. In our case we could verify that very few pixels suffered from these artifacts. 
Straightforward post-processing was used in order to replace
the missing pixels by interpolating the surrounding valid pixels (see panel (b) in Fig.\ref{fig3}).
\\In order to estimate the reliability of the wave height data obtained with the Kinect, we compared it with the measurements from capacitive probes. Here we show two different time series with the same wave forcing period ($1.1$ s) but different amplitudes.
\\For this comparison we placed the capacitive probe at the center of the tank,
one centimeter out of the FoV and upstream from it. The depth measured with the
wave gauge was compared with the output from a single pixel, 
approximately 1.2 cm
(10 pixels) inside the FoV and directly downstream from the capacity probe.
As one can see from Fig.\ref{fig4} we found a good agreement between
amplitude values referred to the two different kind of acquisition: even
water waves with small amplitudes (panel (b) in Fig.\ref{fig4}), although presenting a bigger noise, are
well captured and represented by the device.
Finally, as a proof of concept, 
we attempted to reconstruct the dispersion relation of the recorded wave field.
We explored the sensor's capabilities to reconstruct the dispersion relation connecting the wave number \textit{k} of water waves with their angular frequency $\omega$. According to the linear theory, for arbitrary water depth $h$, such relation is:
\begin{equation} \label{reldisp}
\omega= \sqrt{g\,k\, \tanh(kh)},
\end{equation}

\begin{figure}[h!]
\centering
\includegraphics[width=1\linewidth]{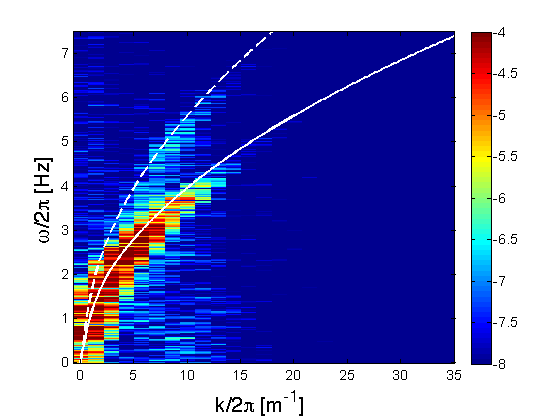}
\caption{Image of space-time Fourier wave spectrum computed on Kinect post processed data. The continuous line represents the relation between wave number $k$ and angular frequency $\omega$ given by the theoretical dispersion law, while the dashed line is its second harmonic. Energy is color coded on a logarithmic scale.}
\label{fig5}
\end{figure}

Were $g$ is the gravity acceleration and $k$ is the wave number. 
\\Clearly this relation requires data points at several frequencies. 
For this analysis we injected energy in the tank with a random forcing and we performed 
a two dimensional space-time Fourier transform of the wave fields. 
We expect the propagation of linear gravity waves to produce intense peaks in the Fourier transform 
corresponding to the $k$-$\omega$ pairs satisfying relation\~(\ref{reldisp}).
\\As one can see from figure \ref{fig5}, we found a high correlation between 
the theoretical dispersion relation and the collocation in $(k,\omega)$ Fourier space of the most energetic modes.

\section{Conclusions\label{conclusions}}
We presented some results on the use for scientific purposes of a commodity
depth sensor, the Microsoft Kinect, originally designed for videogames use. 
The particular application we proposed is the 
measurement of surface waves in water. Other uses of the same sensor have been
proposed, among which the automatic analysis of crowds
\cite{Corbetta2017}. For such application the accuracy of the depth measurement
is less critical since the sensor is used to provide data which can be fed to
shape recognition software.
For the measurement of water waves, the main drawback is the necessity to
render the water opaque. While we showed that this can be done rather
inexpensively using water-soluble paint, further characterization of the
solution is needed in order to clarify the rheological effects of the specific
substance employed.
Our results show that the Kinect can be effectively used for the analysis of waves with amplitudes exceeding a few millimeters, wavelengths between 5 and 40 cm and frequencies up to 6 Hz. Our setup allowed us to measure the dispersion relation of surface waves, thus serving as a proof of concept for the use of the device for non-trivial, multifrequency statistical analysis of a wave-field. All post-processing used
in our analysis can be easily automated, thus making for a cost-effective
detector for surface waves. Further development in the post-processing package can potentially increase the vertical resolution by optimized noise reduction.

\section{Acknowledgments}
Experiments were supported by the European Community Framework Program 7,
European High Performance Infrastructures in Turbulence (EuHIT), Contract No.
312778. Support from the "Departments of Excellence“ (L. 232/2016) grant,
funded by the Italian Ministry of Education, University and Research (MIUR) is
acknowledged. M. O. has been funded by Progetto di Ricerca d'Ateneo CSTO160004.
We thank A. Corbetta and F. Toschi for help in developing the software
interface for Kinect.

\section*{References}

\bibliography{biblio}

\end{document}